\documentclass[usenatbib]{mn2e}
\usepackage{graphicx}

\title{45 Years of Rotation of the Crab pulsar}

\author[Lyne et al.]
	{A.~G.~Lyne$^{1,}$\thanks{E-mail: andrew.lyne@manchester.ac.uk}, C.~A.~Jordan$^1$, 
	F.~Graham-Smith$^1$, 
\newauthor
        C.~M.~Espinoza$^{1,2}$, B.~W.~Stappers$^1$ and P.~Weltevrede$^1$
\\
	$^1$ Jodrell Bank Centre for Astrophysics, School of Physics and Astronomy,
	The University of Manchester, Manchester M13 9PL, UK
\\
	$^2$ Instituto de Astrof\'isica, Facultad de F\'isica,
	Pontificia Universidad Cat\'olica de Chile, Casilla 306,
	Santiago 22, Chile
\\}

\begin{document}

\date{}
\pagerange{\pageref{firstpage}--\pageref{lastpage}} \pubyear{2014}
\maketitle

\label{firstpage}
\begin{abstract}
The 30-Hz rotation rate of the Crab pulsar has been monitored at
Jodrell Bank Observatory since 1984 and by other observatories before
then.  Since 1968, the rotation rate has decreased by about $0.5$\,Hz,
interrupted only by sporadic and small spin up events (glitches).  24
of these events have been observed, including a significant
concentration of 15 occurring over an interval of 11 years following
MJD 50000.  The monotonic decrease of the slowdown rate is partially
reversed at glitches.  This reversal comprises a step and an
asymptotic exponential with a 320-day time constant, as
determined in the three best-isolated glitches.  The cumulative
effect of all glitches is to reduce the decrease in slowdown rate by
about 6\%. Overall, a low mean braking index of $2.342(1)$ is measured
for the whole period, compared with values close to $2.5$ in intervals
between glitches.  Removing the effects of individual glitches reveals
an underlying power law slowdown with the same braking index of 2.5.
We interpret this value in terms of a braking torque due to a dipolar
magnetic field in which the inclination angle between the dipole and
rotation axes is increasing.  There may also be further effects due to
a monopolar particle wind or infalling supernova debris.
\end{abstract}

\begin{keywords}
stars: neutron -- pulsars: general
\end{keywords}

\section{Introduction}

The slowdown of a rotating neutron star is usually understood to
arise from the loss of kinetic rotational energy in the form of
electromagnetic radiation from a rotating dipolar magnetic field
attached to the star.  In this simple model, the slowdown rate of a
pulsar in vacuo, rotating at frequency $\nu$, having a magnetic dipole
moment $M$, moment of inertia $I$ and inclination angle $\alpha$
between the magnetic and rotation axes is given by
\begin{equation} 
\label{cubelaw}
\dot\nu = -\frac{8\pi^2}{3c^3}\frac{M^2 \sin^2\alpha}{I}\nu^3 .
\end{equation}
If $M$, $I$ and $\alpha$ are constant, we would expect to observe a
dependency of $\dot\nu \propto \nu^3$ as the pulsar slows down.
However, other braking mechanisms are possible with different
dependence upon $\nu$ and, in a more general power-law slowdown of the
form
\begin{equation}
\label{kappa}
\dot\nu=-\kappa\nu^n  ,
\end{equation}

\noindent
the value of the exponent of $\nu$ is known as the braking index $n$,
being 3 for dipolar magnetic torque.  Other mechanisms have been
proposed which would give lower values of $n$; for example,
\citet{mt69} showed that the torque from a simple outflow of particles
would lead to $\dot\nu \propto \nu^1$, i.e. $n=1$.  The value of $n$ may also
differ from 3 if, for instance, the values of $M$, $\alpha$ or $I$ in
equation~(\ref{cubelaw}) vary with time \citep{br88}.  In order to
explore the physics of this slowdown, the value of $n$ can in
principle be determined from observation of higher-order frequency
derivatives; differentiation of equation~(\ref{kappa}) shows that the
second and third derivatives should be related to $n$ by
\begin{eqnarray}
\label{ndef}
n&=& {\ddot\nu\nu\over \dot\nu^2} \;\;\;\; {\rm and}\\
\label{F3}
n(2n-1)&=&{\stackrel{...}{\nu}\nu^2\over \dot\nu^3}.
\end{eqnarray}

For most pulsars, evaluation of the rotational slowdown law is
confined to measurement of the rotational frequency $\nu$ and its
first derivative $\dot\nu$. For old pulsars, the second derivative
$\ddot\nu$ expected from equation~(\ref{ndef}) is usually unmeasurably small,
while in many younger pulsars any secular behaviour is often confused
by unpredictable changes in rotation rate, in the form of either
timing noise or glitches.  Timing noise is seen as slow quasi-random
changes in rotation rate and arises from magnetospheric instabilities
\citep{lhk+10}, while glitches are almost instantaneous increases in
rotation rate, often followed by some associated transient behaviour
and have their origin in the neutron star interior \citep{elsk11}.

Because of these effects, values of braking index have been reliably
established for only eight pulsars.  For the Crab pulsar, $\ddot\nu$
has been measured between glitches \citep{lps93}, leading to an
observed value $n_{\rm obs}=2.51(1)$, significantly less than the
value of $n=3$ expected for braking by magnetic dipole radiation.  The
same is true for all the other seven pulsars (Table \ref{indices}).
These results indicate that the physical process causing the slowdown is not 
just simple dipolar electromagnetic radiation.

\begin{table} 
\caption{Measured braking indices for young pulsars} 
\label{indices} 
\begin{tabular}{llll} 
\hline\hline 
PSR & $n$ & Reference \\ 
\hline 
B0531+21(Crab)& 2.51(1) & \citet{lps93} \\
B0540$-$69      & 2.14(1) & \citet{lkg+07} \\ 
B0833$-$45(Vela)& 1.4(2)  & \citet{lpgc96} \\ 
J1119$-$6127    & 2.684(2)& \citet{wje11} \\ 
B1509$-$58      & 2.839(1)& \citet{lkg+07} \\ 
J1734$-$3333    & 0.9(2)  & \citet{elk+11}\\
J1833$-$1034    & 1.857(1)& \citet{rgl12}\\ 
J1846$-$0258    & 2.65(1) & \citet{lkg+07} \\ 
\hline 
\end{tabular} 
\end{table}

\begin{figure}
 \includegraphics[width=8.5cm]{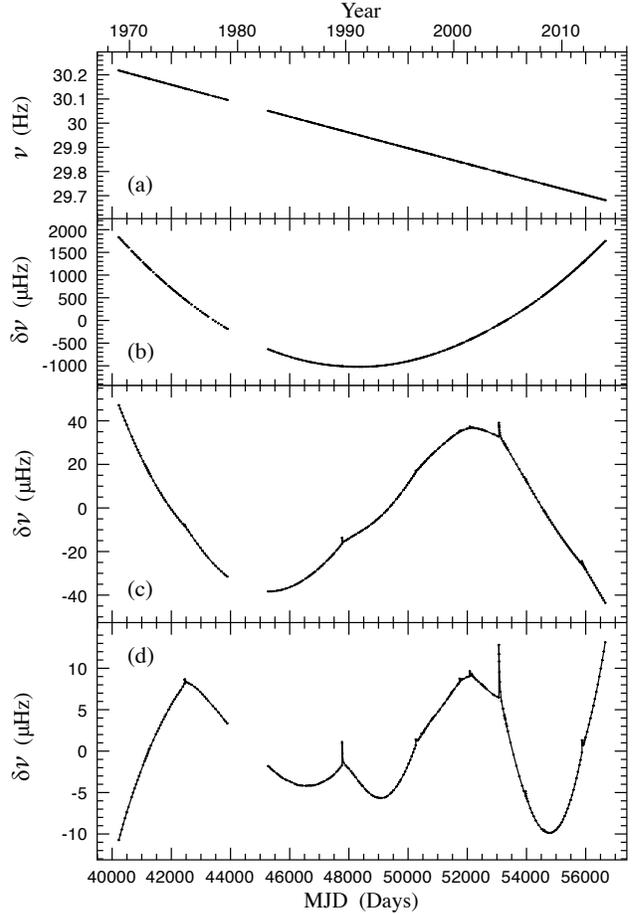}
 \caption{The spin-frequency history of the Crab pulsar over 45 years.  
   (a) The observed spin-frequency $\nu$ determined from fits to 
    100-day data sets every 50 days, showing the monotonic slow-down 
    of the pulsar.
   (b), (c) and (d) The frequency residuals $\delta\nu$ after fitting 
    to the values in (a) simple slow-down models involving frequency and
    respectively one, two and three spin-frequency derivatives in the 
    Taylor Series of equation~(\ref{taylor}).  The fitted values of 
    $\nu_0$, $\dot{\nu}_0$, $\ddot{\nu}_0$ and $\stackrel{...}{\nu}_0$ 
    for (d) are given in Table~\ref{spinpars}.}
 \label{nu}
\end{figure}

\begin{figure}
 \includegraphics[width=8.5cm]{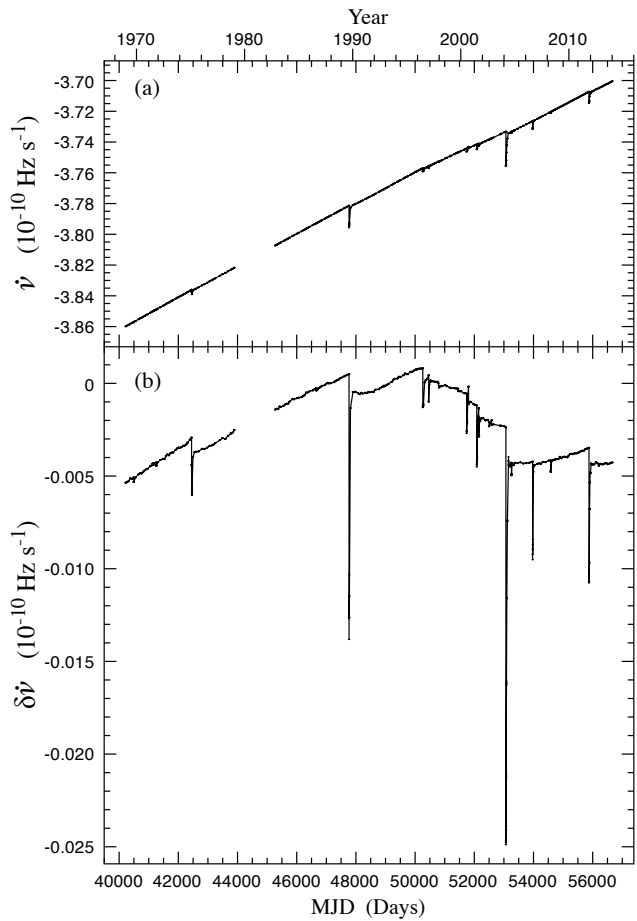} 
 \caption{The slowdown rate $\dot\nu$ of the Crab pulsar over 45 years.  
   (a) Observed values of the first derivative $\dot\nu$ determined from
    fits to 100-day data sets every 50 days.  The magnitude of the 
    slowdown rate $\vert\dot\nu\vert$ is decreasing overall, with small 
    step increases at the glitches.  
   (b) Frequency derivative residuals $\delta\dot\nu$ on an expanded 
    scale, obtained by subtracting from (a) a linear model using the values of
    the first two frequency derivatives given in Table~\ref{spinpars}.}
 \label{nudot}
\end{figure}

In this paper we report on the measurement and analysis of the
rotation rate of the Crab pulsar from 1968 to 2013.  This 45-year
time-baseline amounts to about 5\% of the pulsar lifetime and allows
the spin-down of the Crab pulsar to be described over a period which
includes many glitches and provides more details of the cumulative
effect that they have on the long-term spin-down \citep{lps93,sj03}.
Elsewhere, the same data have been used to examine the statistics and
physical details of the glitches \citep{eas+14} and to study the
evolution of the radio pulse emission over this time \citep{lgw+13}
to enable a comprehensive picture of the evolution of the pulsar.

\section{Observations and Basic Analysis}

The rotation of the Crab pulsar has been monitored by daily
observations at Jodrell Bank Observatory since 1984, mainly using the
13-m radio telescope at 610 MHz \citep{lps88,lps93}.  Regular
observations with the 76-m Lovell telescope at around 1400-1700 MHz,
designed to monitor any changes in dispersion measure, also contribute
to the dataset.

These data have been supplemented with earlier observations taken at
Arecibo \citep{girp77} and in the optical at Princeton
\citep{gro75a} and Hamburg \citep{loh81}.
There are no observations available between February 1979 and
February 1982, this being the only significant gap with no data.
There are in total appproximately 11,000 times of arrival (TOAs) and
together the measurements comprise a record of the rotation of the
pulsar over a total of 45 years from November 1968 to December 2013.

In order to study the long-term rotational history of the pulsar, we
have used standard procedures to reduce the TOAs to the barycentre of
the Solar System.  We have then fitted values of the rotation
frequency $\nu$ and its first two derivatives $\dot\nu$ and $\ddot\nu$
over time spans of approximately 100 days.  Such analyses were
repeated with the central reference time advancing by typically 50
days between analyses.  Close to glitches, the time spans were
adjusted in such a way that no analysis was performed over a glitch,
so that one analysis ended and another started close to the epoch of
the glitch.

These time sequences of rotational frequencies and first derivatives
provide the main forms of the data that we use to study the long-term
behaviour of the pulsar in this paper.  Figs.~\ref{nu}a and
\ref{nudot}a illustrate the evolution of the rotational frequency
$\nu({\rm t})$ and slowdown rate $\dot\nu({\rm t})$ over the 45 years.
The rotational slowdown of the pulsar is evident in Fig.~\ref{nu}a,
falling by about 0.5~Hz during this time.   
The slowdown rate (Fig.~\ref{nudot}a) also shows a general reduction in
magnitude with time, but there are also considerable long-term effects
resulting from glitches, which we investigate further in a later
section.

Following convention, it is instructive to characterise the variation
in rotation frequency with time as a Taylor series of derivatives of
the form:
\begin{equation}
\nu(t)= \nu_0 +\dot\nu_0(t-t_0)+{1\over 2}\ddot\nu_0(t-t_0)^2 
            + {1\over 6}\stackrel{...}{\nu}_0(t-t_0)t^3+\delta\nu(t).
\label{taylor}
\end{equation}
By fitting such a model to the frequency data, we evaluate $\nu_0$ and
the first three derivatives, leaving a residual $\delta\nu(t)$.  In
Figs.~\ref{nu}b-\ref{nu}d we see the residuals resulting from fitting
and subtracting Taylor series consisting of 1, 2 and 3 derivative
terms, respectively. In each case, the residuals are dominated by the
next, unfitted derivative term.  The remaining residuals shown in
Fig.~\ref{nu}d, which might in previous analyses have been designated
as timing noise, are related to glitch activity, which we discuss
later.  The best fit values obtained from the 3-derivative fit are
presented in Table~\ref{spinpars}. These parameters represent the
long-term behaviour of the pulsar, including the effects of glitches.

Using equation~(\ref{ndef}) and the values in Table~\ref{spinpars}, a
braking index $n=2.349(1)$ is obtained.  Using this braking index in
equation~(\ref{F3}) it is found that the expected value of
$\stackrel{...}{\nu}$ is $-0.52\times10^{-30}$~Hz\,s$^{-3}$, which is
about one fifth of the measured value.  This shows that the general
long-term slowdown of the Crab pulsar is not a simple power-law. We
show later that removal of the effects of glitches brings the slowdown
close to a simple power-law.

Extrapolating the Taylor series back in time to investigate the
behaviour in earlier years should evidently be approached with
caution, but we remark that the series indicates an original rotation
rate of 58~Hz (a period of 17 milliseconds) at birth in AD~1054.

\begin{table}
 \caption{Observed and derived parameters for the Crab pulsar.  The
 spin parameters were obtained by fitting a 3-derivative Taylor series
 to the frequency data presented in Fig.~\ref{nu}a.  The standard
 errors determined in the analysis are given in parentheses after the
 values and are in units of the least significant digit.}
 \label{spinpars}
 \begin{tabular}{ll}
 \hline \hline
 \multicolumn{1}{c}{Parameter} & \multicolumn{1}{c}{Value}	    \\
 \hline
 Data span (MJD)                                    & 40175$-$56665 \\
 Epoch t$_0$(MJD)				    & 48442.5       \\
 $\nu_0$ (Hz)					    & 29.946923(1) \\
 $\dot{\nu}_0$ ($10^{-10}$~Hz\,s$^{-1}$)	    & $-$3.77535(2) \\
 $\ddot{\nu}_0$	($10^{-20}$~Hz\,s$^{-2}$)	    & 1.1147(5)     \\
 $\stackrel{...}{\nu}_0$ ($10^{-30}$~Hz\,s$^{-3}$)  & $-$2.73(4)    \\
 RMS residual (Hz)                                  & 0.0000058     \\
 Maximum frequency residual (Hz)                    & 0.000013      \\
 Mean braking index $n$                             & 2.342(1)      \\
 \hline
 \end{tabular}
\end{table}

\section{The 24 glitches}

The monotonic rotational slowdown of the Crab pulsar seen in
Fig.~\ref{nu} is occasionally reversed discontinuously at a glitch, in
which $\nu$ increases by a small step $\Delta\nu$ of order
$10^{-9}\nu$ to $10^{-7}\nu$, followed by a nearly complete recovery
within about 20 days of the event.  The magnitude of the slowdown rate
$\vert\dot\nu\vert$ (Fig.~\ref{nudot}) also presents an initial step
increase at a glitch, again partially reversing the long-term trend
and showing the corresponding short-term recovery \citep{lps93,wbl01}.
However, in contrast with the relaxation in $\nu$, the recovery in
slowdown rate is not complete and a persistent step
($\Delta\dot\nu_{\rm p}$) is commonly observed after glitches.  First
remarked upon by \citet{girp77} and \citet{dp83} in relation to the
glitch of 1975, these persistent steps are a general feature of
glitches in this pulsar and have an appreciable effect on the overall
slowdown behaviour.

\begin{figure}
\includegraphics[width=8.5cm]{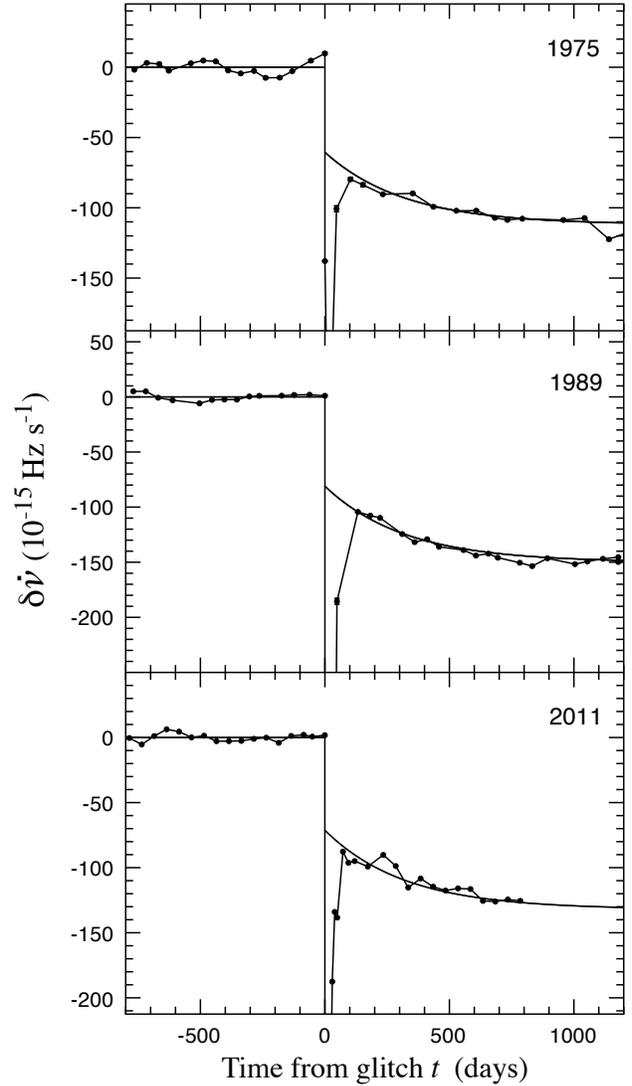}
\caption{The variation in slowdown rate $\dot\nu$ of the Crab pulsar
near to the three large isolated glitches which occurred in 1975, 1989
and 2011.  In each case, the frequency derivative residuals
$\delta\dot\nu$ were obtained relative to a linear model involving the
first two frequency derivatives fitted to $\dot\nu$ over the 800 days
preceding the glitch, which occurred at day 0.  Each glitch is
followed instantly by a large negative transient (increase in
magnitude) in $\dot\nu$. The main transient behaviour ceases after
about 100 days, revealing a persistent negative offset which continues to
increase in a quasi-exponential manner on a
timescale of around 320 days. The smooth lines are the fits to the
post-glitch data described by equation \ref{vdot_pers}.
}
\label{3glits}
\end{figure}

Unfortunately, the study of the long-term impact of glitches is
often contaminated by the occurrence of other, nearby glitches.
However, there are three large glitches which are ``isolated'', each
having no other detectable glitches within 800 days before or 1200
days after the epoch of the glitch.  These three glitches, in 1975,
1989 and 2011 allow the nature of these persistent steps to be
demonstrated in Fig.~\ref{3glits}.  As pointed out by \citet{lps93} in
relation to the glitch of 1989, the persistent step increase in
slowdown rate consists of an instantaneous component followed by a
further quasi-exponential asymptotic increase on a timescale of about
265 days.  We now see in Fig.~\ref{3glits} that the same description
applies closely to all three large, isolated glitches: all three step
changes in $\dot\nu$ have different amplitudes and all three show a
further increase in $\dot\nu$ which grows exponentially for at least a
year after the initial step.  Additionally, the relative amplitudes of
the step and the exponential and the time scales of the exponential
are identical to within the fitted errors for all three. As a result,
we conclude that the long-term recovery can be represented by a
single-parameter function of the form:

\begin{equation}
  \delta\dot{\nu} = \left\{ \begin{array}{ll} 
   0                                                           & \mbox{if $t<0$} \\
   \Delta\dot{\nu}_{\rm p} \times (0.46\times {\rm exp}(-t/320)-1.0) & \mbox{if $t>0$,}
                             \end{array}
                     \right. 
\label{vdot_pers}
\end{equation}
where $t$ is the time from the glitch epoch in days and
$\Delta\dot{\nu}_{\rm p}$ is the total persistent step in slowdown
rate.  Note that the asymptotic exponential is in the same sense as
the initial step in $\dot\nu$, unlike the short-term recovery at a
glitch.

We now regard this as an intrinsic component of all glitches, although
for many it is often obscured by the recovery from previous glitches
or the occurrence of later glitches. The asymptotic exponential
component has a timescale of 320$\pm$20 days and comprises 46\%, or
almost half, of the total increase of slowdown rate.  In what follows,
we assume that these values apply to all glitches in the Crab pulsar.

\begin{figure}
\includegraphics[width=8cm]{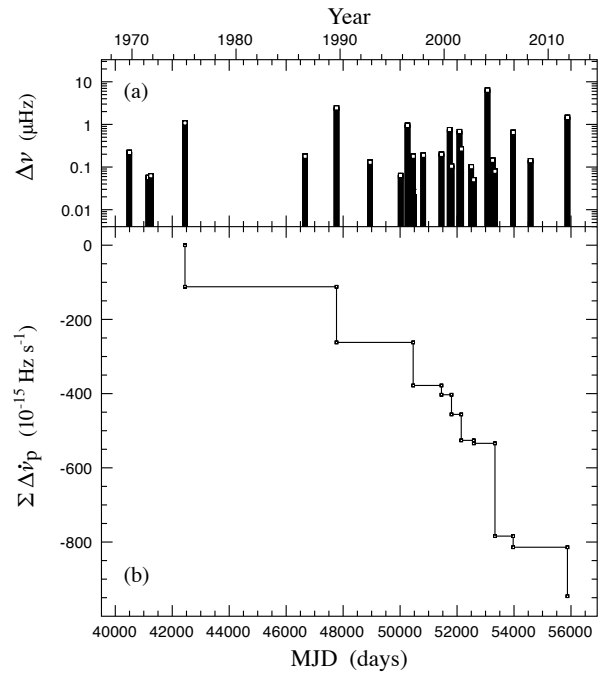}
\caption{Glitches in the Crab pulsar. 
 (a) The distribution in time and magnitude of the steps $\Delta\nu$ in
 spin-frequency of the 24 glitches in Table \ref{glitch_list}. 
 (b) The cumulative effect of the persistent steps in slowdown rate 
$\Delta{\dot\nu}_{\rm p}$ at the glitches.}
\label{glitches}
\end{figure}

\begin{figure}
\includegraphics[width=8cm]{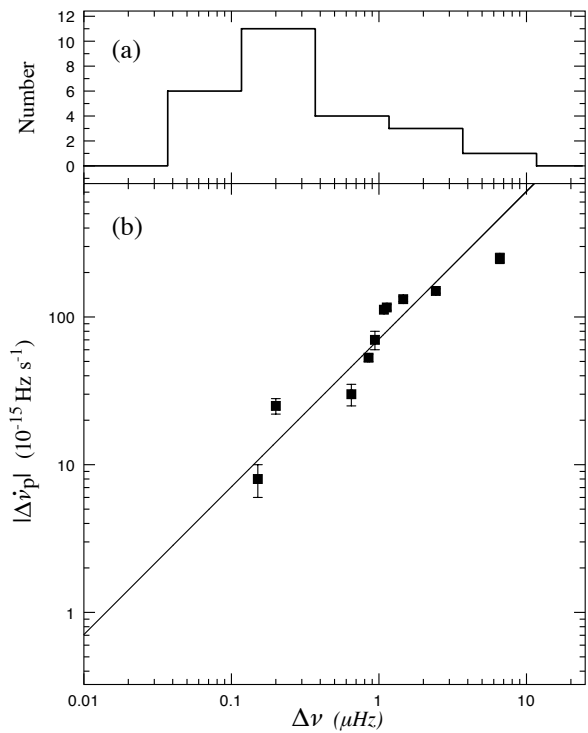}
 \caption{Glitch sizes.
 (a) Histogram of frequency steps $\Delta\nu$ on a logarithmic scale
   for the 24 glitches in Table \ref{glitch_list}.
 (b) The inter-dependence of the persistent change in slow-down rate
$|\Delta\dot{\nu}_{\rm p}|$ and the step-change in rotation frequency
$\Delta\nu$ for 10 glitches. The diagonal line has a slope of unity. 
Only events for which $\Delta\dot{\nu}_{\rm p}<0$ are shown.}
\label{dfdf1p}
\end{figure}

Table~\ref{glitch_list} lists 24 glitches, giving the date and MJD of
occurrence, the step change $\Delta\nu$, the fractional value $\Delta
\nu/\nu$ and the persistent change of slowdown rate $\Delta \dot
\nu_{\rm p}$.  This is a robust list which includes all glitches for
which $\Delta\nu> 0.01 \mu$Hz.  Any glitches smaller than this are of
similar size to the period variations produced by the noise present in
the data.  Any glitch which may have occurred during the gap in
observations in 1981 left no obvious change in slowdown rate, and
would presumably have been small and would not affect the results of
our analysis.  Glitch epochs and frequency step sizes were taken from
\citet{elsk11} and \citet{ejb+11}.  The sequence of glitches is
illustrated in Fig.~\ref{glitches}a, which shows the date of
occurrence and the size $\Delta\nu$ of each glitch (on a logarithmic
scale). Note particularly the pattern of substantially increased
activity from MJD $\sim 50000$ to MJD $\sim 54000$ (years 1995-2006).
A histogram of glitch sizes $\Delta\nu$ is shown in
Fig.~\ref{dfdf1p}a.  Espinoza et al. (2014)\nocite{eas+14} have
analysed the Jodrell Bank data on the Crab pulsar and conclude
that the decrease in numbers of glitches towards smaller sizes is
intrinsic and is not related to the detection capabilities.

The persistent steps in frequency derivative were estimated by fitting
a function of the form given by equation~(\ref{vdot_pers}) to the observed
$\dot{\nu}$ data centred on the glitch epoch.  In order to avoid
contamination by short-term glitch transient recoveries, data taken
within 90 days following each glitch were not used.  In some cases the
step is very small and no significant measurement was possible. In
other cases, because of the presence of other nearby glitches, the
available data points were insufficient to perform the measurement,
although in a few cases it was possible to measure the combined effect of 
two or three closely-separated glitches (Table \ref{glitch_list}). 
Without being extremely precise, this method effectively measures the
basic trends we are studying and it is not affected by the necessary
assumptions of more complicated models.  Our values are roughly
consistent with existent measurements using such models
\citep[e.g.][]{wbl01,wwty12}.

The step in $\Delta\dot{\nu}_{\rm p}$ at a glitch is loosely related to
the step-change in rotation frequency $\Delta\nu$.  The relationship
is shown in Fig.~\ref{dfdf1p}b, which includes data from 15 glitches
(some points combine data from two or three adjacent glitches).  The
approximately linear relationship is fitted by the diagonal line
of unity slope, given by

\begin{equation}
|\Delta\dot{\nu}_{\rm p}| = 7\times10^{-8}\times\Delta\nu \;\;\rm{ Hz\; s^{-1}},
\label{dfdf1pe}
\end{equation}
where $\Delta\nu$ is in Hz.

The persistent slowdown increase is usually attributed to a reduction
in effective moment of inertia due to pinning of neutron superfluid
vortices to other internal components.  Because a rotating superfluid
slows down only if its vortices are allowed to move apart, the pinning
of some vortices decouples a fraction of the superfluid from the rest
of the star and reduces the effective moment of inertia of the
rotating star.  Equation~(\ref{dfdf1pe}) indicates that the re-pinning
of vortices is more effective after larger glitches.  

We offer no explanation for the surprising phenomenon of the long-term
asymptotic increase in $\dot\nu$ and continue to view all increases of
slowdown rate at glitches as decreases in effective moment of inertia,
presumably due to vortex pinning.  We note that if there is no
relaxation in this cumulative pinning, the proportional change in
moment of inertia has amounted to $0.3\%$ in 45 years.  We remark that
such a large rate of change cannot persist for more than a few
thousand years, by which time a large proportion of the effective
moment of inertia of the neutron star would have disappeared.

\begin{table}
\caption{The steps in rotation rate and slowdown rate for the 24
glitches observed between 1968 and 2013. This table is derived from
\citet{eas+14}}.
\label{glitch_list}
\begin{tabular}{llclc} 
\hline\hline
\multicolumn{1}{c}{Date} &
\multicolumn{1}{c}{MJD} &
\multicolumn{1}{c}{$\frac{\Delta\nu}{\nu}$} &
\multicolumn{1}{c}{$\Delta\nu$} &
\multicolumn{1}{c}{$\Delta\dot{\nu}_p$} \\
\multicolumn{1}{c}{ } &
\multicolumn{1}{c}{days} &
\multicolumn{1}{c}{$10^{-9}$} &
\multicolumn{1}{c}{$\mu$Hz} &
\multicolumn{1}{c}{$10^{-15} \textrm{s}^{-2}$} \\
\hline 
1969 Sep & 40491.84(3)  & 7.2(4)  & 0.22(1)  & $-$          \\    
1971 Jul & 41161.98(4)  & 1.9(1)  & 0.057(4) & $-$          \\    
1971 Oct & 41250.32(1)  & 2.1(1)  & 0.062(3) & $-$          \\    
1975 Feb & 42447.26(4)  & 35.7(3) & 1.08(1)  & $-$112(2)    \\ 
1986 Aug & 46663.69(3)  & 6(1)    & 0.18(2)  & $-$          \\     
\\ 				            
1989 Aug & 47767.504(3) & 81.0(4) & 2.43(1)  & $-$150(5)    \\  
1992 Nov & 48945.6(1)   & 4.2(2)  & 0.13(1)  & $-$          \\ 
1995 Oct & 50020.04(2)  & 2.1(1)  & 0.063(2) & $-$          \\     
1996 Jun & 50260.031(4) & 31.9(1) & 0.953(4) & $^*$         \\ 
1997 Jan & 50458.94(3)  & 6.1(4)  & 0.18(1)  & $-116(5)^\dagger$  \\ 
\\ 				            
1997 Dec & 50812.59(1)  & 6.2(2)  & 0.19(1)  & $-$          \\   
1999 Oct & 51452.02(1)  & 6.8(2)  & 0.20(1)  & $-$25(3)     \\   
2000 Jul & 51740.656(2) & 25.1(3) & 0.75(1)  & $^*$         \\ 
2000 Sep & 51804.75(2)  & 3.5(1)  & 0.105(3) & $-53(3)^\dagger$   \\   
2001 Jun & 52084.072(1) & 22.6(1) & 0.675(3) & $^*$         \\
\\ 				            
2001 Oct & 52146.7580(3)& 8.87(5) & 0.265(1) & $-70(10)^\dagger$  \\ 
2002 Aug & 52498.257(2) & 3.4(1)  & 0.101(2) & $^*$            \\ 
2002 Sep & 52587.20(1)  & 1.7(1)  & 0.050(3) & $-8(2)^\dagger$    \\   
2004 Mar & 53067.0780(2)& 214(1)  & 6.37(2)  & $^{*}$      \\ 
2004 Sep & 53254.109(2) & 4.9(1)  & 0.145(3) & $^{*}$      \\ 
\\ 				            
2004 Nov & 53331.17(1)  & 2.8(2)  & 0.08(1)  & $-250(20)^{\dagger\dagger}$ \\ 
2006 Aug & 53970.1900(3)& 21.8(2) & 0.65(1)  & $-$30(5)     \\
2008 Apr & 54580.38(1)  & 4.7(1)  & 0.140(4) & $-$          \\
2011 Nov & 55875.5(1)   & 49.2(3) & 1.46(1)  & $-$132(5)    \\
\hline
\end{tabular}
$^\dagger$incorporates the persistent step of the previous glitch.
$^{\dagger\dagger}$incorporates the persistent steps of the previous two glitches.
\end{table}

\citet{mpw08} include the Crab pulsar glitches in a comprehensive
analysis of the statistics of the distributions of size and intervals
between glitches, testing the hypothesis that these are determined by
a random process of self-organised criticality.  The incidence of
glitches in most pulsars appears to be random, and \citet{wwty12} show
that the distribution of interval times in the Crab pulsar is
Poissonian, although some pulsars show a quasi-periodicity
\citep{lel99,mpw08,wwty12}.

\begin{figure}
\includegraphics[width=8cm]{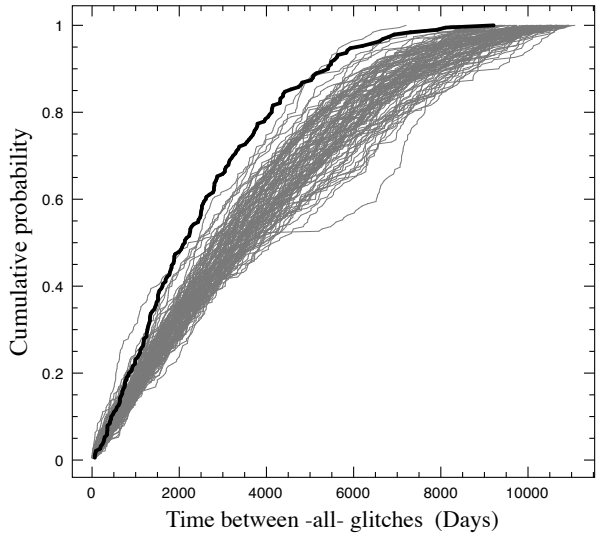}
 \caption{The cumulative distribution of intervals between all 20
 glitches occurring between MJD 46000 and MJD 56600, as observed (bold
 line) and in 100 random simulations. }
\label{P_cumul}
\end{figure}

We nevertheless draw attention to the cluster of 15 glitches between
MJD 50000 and 54000 seen in Fig.~4, which suggests that there may be
an extra complexity in the system.  We require a statistical test for
the hypothesis that this cluster was a chance concentration of
glitches which occurred at random.  The \citet{wwty12} test was not
sensitive to clustering, which is better revealed by the distribution
of intervals between all glitches, rather than only consecutive ones.  We
therefore compare the observed distribution of intervals between all
glitches with those from simulated glitch sequences and show that the
probability that the observed distribution is random is low.

We confined our analysis to the 20 glitches which occurred in the last
30 years since high-cadence monitoring commenced in January 1984. As
explained in \citet{eas+14}, the data in this period constitute a
uniform set and are complete, indicating a mean rate of 0.67
glitches/yr. The bold line in Figure \ref{P_cumul} shows the
cumulative distribution of all 190 glitch separation times in this set
of data.  We then generated 10,000 simulated distributions of 20
events, each occurring at a random time within a 30-year period;
in Fig.~\ref{P_cumul}, we show 100 of these realisations (randomly
picked) for comparison with the observed distribution.

The curve derived from the observed data lies on the outside of the
spread of curves from the simulations, indicating a low probability
that the observed cluster was a random occurrence. In particular,
there is an excess of small separation times in the observed
cumulative distribution, which is therefore steeper than most of the
simulated distributions. This indicates that the glitches are more
clustered than can be expected from a random occurrence of
glitches. We also compared the means of the separation times in the
distributions; the mean of the 190 observed glitch separation times was
less than the mean in all except 52 (~0.5\%) of the 10,000 simulated
distributions. We therefore conclude that it is unlikely that the
cluster of glitches is a statistical accident. We note that the
11-year period of the cluster coincides with a period of an anomaly in
the underlying slowdown rate, which is seen in Fig.~\ref{underlying_n}
and will be discussed in the next Section.

\section{The effect of glitches on the slowdown}

Fig.~\ref{glitches}b shows the cumulative effect of the steps
$\Delta\dot{\nu}_{\rm p}$ at the 17 glitches for which this quantity
was measured.  The total cumulative effect over 45 years is an
increase in slowdown rate $\vert\Sigma\Delta\dot\nu_{\rm p}\vert$ of
$946(25)\times10^{-15}$~Hz\,\,s$^{-1}$.  The small but appreciable
overall negative contribution made by the glitches to the second
derivative $\ddot\nu$ may be evaluated by dividing the total sum by
the time interval, giving a value of
$-0.066(4)\times10^{-20}$~Hz\,\,s$^{-2}$, which is approximately 6\%
of the second derivative $\ddot\nu$ evaluated from the long-term
analysis (Table~\ref{spinpars}).  

\begin{figure}
\includegraphics[width=8.5cm]{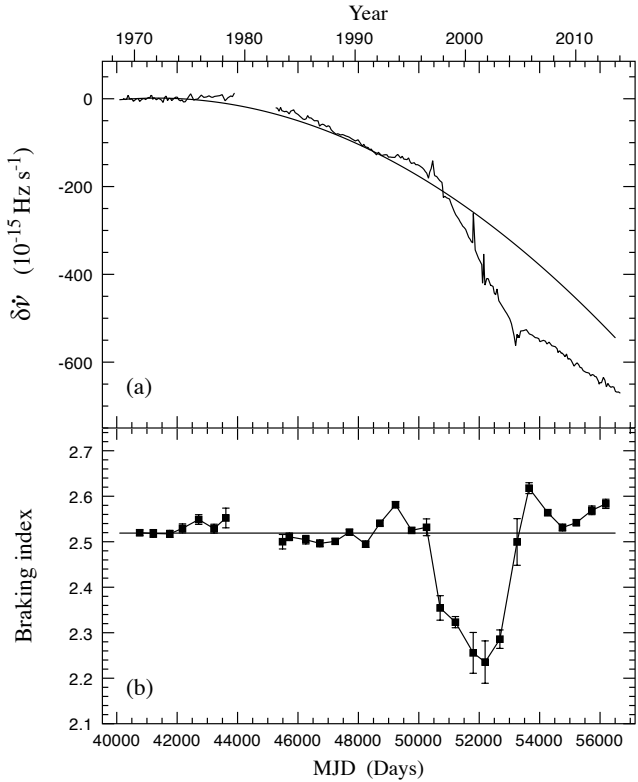}
\caption{The underlying slowdown of the Crab pulsar after removal of
the effects of the glitches.  a) The frequency derivative residuals
$\delta\dot\nu$ obtained by subtracting from $\dot\nu$ data the
persistent steps $\Delta\dot\nu_{\rm p}$ at the glitches given in
Table~\ref{glitch_list}, removing short-term glitch transients by
excluding data within 90 days following each glitch, and by removing
the main linear trend as measured from the start to MJD~50458. The
smooth curve is the expected behaviour during this era for a constant
value of $n=2.519$, showing a curvature corresponding to a value of
$\stackrel{...}{\nu} = -0.636\times 10^{-30}$ Hz\,s$^{-3}$ as
calculated from equation~(\ref{F3}).  b) The underlying braking index
of the Crab pulsar using equation~\ref{ndef}, evaluated over sections
of 1000 days at 500-day intervals.  The values of $\nu$ and
$\dot{\nu}$ are the mean values of the data in each interval in
Figs.~\ref{nu}a and \ref{nudot}a, and $\ddot\nu$ is taken from the
slope of the corresponding data in (a).  The horizontal line at
$n=2.519$ corresponds to the curve in (a).}
\label{underlying_n}
\end{figure}

We now address the notion that the slowdown is an underlying steady
phenomenon with superimposed cumulative steps in the slowdown rate
caused by glitches.  If the persistent steps $\Delta\dot{\nu}_{\rm
p}$, as tabulated in Table~\ref{glitch_list} and with the form
described by equation \ref{vdot_pers}, are removed from the
$\dot{\nu}$ data we obtain a smoother, ``corrected'' $\dot{\nu}$
evolution which shows the underlying slowdown.
Fig.~\ref{underlying_n}a shows frequency derivative residuals obtained
by removing from the corrected $\dot{\nu}$ data a new linear 2-term
Taylor series fitted to the restricted 6-year span up to MJD 42447,
this being the epoch of the 1975 glitch.  This fit yields values of
$\dot{\nu}=-3.84796(1)\times10^{-10}$~Hz\,\,s$^{-1}$ and
$\ddot\nu=1.236(1)\times10^{-20}$~Hz\,\,s$^{-2}$ at MJD 41338.67.  We
now find from equation~(\ref{ndef}) that during this span, the braking
index was $n=2.519(2)$, and from equation~(\ref{F3}) the expected
value of $\stackrel{...}{\nu} = -0.636(1)\times 10^{-30}$
Hz\,\,s$^{-3}$.  The smooth curve in Fig.~\ref{underlying_n}a is the
calculated variation in $\dot{\nu}$ including the curvature due to
this term, and in general is seen to track the data well.

However, at around MJD 50000, the slope in Fig.~\ref{underlying_n}a
changes abruptly, recovering at around MJD 54000, after an accumulated
offset in $\dot\nu$, to approximately the expected slope.  This is the
same 11-year span that contains the large concentration of glitches.
The accumulated deviation in this 11-year span amounts to
$\sim-200\times 10^{-15}$ Hz\,\,s$^{-1}$, and is in addition to the
$\sim-522\times 10^{-15}$ Hz\,\,s$^{-1}$ already accounted for by the
observed persistent glitch contributions $\Delta\dot{\nu}_{\rm p}$ in
this period, and may be compared with the total change during this
time of $\sim3500\times 10^{-15}$ Hz\,\,s$^{-1}$ due to the underlying
slowdown.  It seems unlikely that the deviation can be due to glitches
which are below the threshold of our measurements, since
Fig.~\ref{dfdf1p}a shows a deficit of small glitches
\citep{eas+14}. More likely, the accumulated deviation is due to a
phenomenon related to glitches which is not reflected in our present
model.

The variations in $\delta\dot\nu$ may also be interpreted as changes
in the underlying value of the braking index $n$ with time. These
changes are summarised in Fig.~\ref{underlying_n}b, in which the value
of the braking index is close to 2.5 throughout most of the 45 years,
except for the period of high glitch activity from MJD 50000-54000,
when it takes a value of about 2.3.

We remark that the process of separating the effects of glitches from
an underlying steady rotational evolution provides a good description
of the overall behaviour, with braking index $n = 2.5$. The small
effect on the index seen in Fig.~\ref{underlying_n}b remains
unexplained.

\section{The slowdown power law}
\label{powerlaw}

Why is the braking index different from 3, as expected from
equation~(\ref{cubelaw}) for energy loss through electromagnetic
dipole radiation? Either this equation is inadequate because angular
momentum is also lost through a mechanical process such as an
outflowing jet or interaction with an external fall-back disk of
supernova remains \citep{mph01a}, or one or more of the parameters
$I$, $M$, or $\alpha$ of the equation is changing with time. These two
categories might be called the unipolar and dipolar approaches in
which the loss of angular momentum may lead to very different slowdown
laws $\dot\nu\ \propto\nu$ and $\dot\nu\propto\nu^3$ respectively.  As
pointed out by \citet{hck99}, \citet{xq01}, and \citet{cl06}, a
combination of both processes could account for the observed low
braking indices of young pulsars.

\subsection{The wind component}

The Crab pulsar, along with other young pulsars, generates a powerful
particle stream whose effect is observed as a wind nebula
(e.g. \citet{gs06}).  The possible effect of angular momentum carried
away by this particle stream has been much debated, following
\citet{mt69} who pointed out that if the slowdown were predominantly
due to wind, the torque would follow the first rather than the third
power of the rotation rate.  Other values of braking index of less
than three due to particle flows have also been suggested
(e.g. \citet{wxg03}).  An alternative approach by \citet{mph01a}
suggests that interaction between the rotating pulsar and an accretion
disc of infalling supernova remains, forming outside the magnetosphere
but coupled by a propeller torque, would lead to a similar slowdown
law.  \citet{yps12} consider the effect of this process on the timing
properties of old as well as young pulsars.

For instance, if the departure of the braking index from 3 is due
to such a dynamic process with braking index $n=1$, the
slowdown can be represented by a combination of two power laws:
\begin{equation} 
\dot\nu= -(A\nu^3 + B\nu ).
\end{equation}
If at the present time the dynamic component increases the slowdown
rate by a factor $(1+\epsilon)=(1+B/A\nu^2)$, it is easy to show that
the observed braking index is
\begin{equation} 
n={\nu\ddot\nu \over\dot\nu^2}=\frac{\epsilon+3}{\epsilon+1} 
\label{epsilon1}
\end{equation}
and
\begin{equation} 
\epsilon=\frac{3-n}{n-1} .
\label{epsilon2}
\end{equation}

If the wind (or the disk) was to be responsible for the whole of the
difference between $n_{\rm dip}=3$, the expected value from
magnetic-dipole braking, and $n_{\rm obs}$, the observed value, then
equation~(\ref{epsilon2}) shows that for $n_{\rm obs}=2.5$, the wind
accounts for approximately one third of the slowdown torque.  A more
precise modelling of the effect of the wind or disk might lead to a more
accurate assessment of this proportion.  For the Vela pulsar, the very
low value of $n_{\rm obs}=1.4$ suggests that the wind is responsible
for ${4\over 5}$ of the slowdown torque.

\subsection{A changing magnetic dipole}

If alternatively the low braking index is predominantly due to
magnetic dipole radiation in which the coefficient $\kappa$ in
equation~(\ref{kappa}) is a function of time $\kappa(t)$, then the observed
braking index will be 
\begin{equation}
n_{\rm obs}={\ddot\nu\nu \over \dot\nu^2}=n_{\rm dip}+{\dot\kappa(t) \nu \over
\kappa(t)\dot\nu} .
\label{kappadot2}
\end{equation}
The observed interglitch value of braking index for the Crab pulsar,
$n_{\rm obs}= 2.5$, would require
$\dot\kappa/\kappa\simeq2.0\times10^{-4}$ yr$^{-1}$.  Since
\begin{equation} 
\kappa \,\, {\propto} \,\, M^2 \sin^2\alpha I^{-1},
\label{coefficient}
\end{equation}
we evaluate the changes in these three parameters which
might account for the low value of the braking index.

From equation~(\ref{kappadot2}) the effects are given by
\begin{equation}
n_{\rm obs}= n_{\rm dip}+{\nu\over\dot\nu} \left(-{{\dot I} \over
I}+2{\dot\alpha \over \tan\alpha}+2{\dot M \over M}\right).
\label {bdot}
\end{equation}

The overall changes in the Crab slowdown are too large to be explained
by simple changes in ellipticity and the consequent changes in the
moment of inertia $I$ of a rotating ellipsoid. An apparent reduction
in $I$ might however be due to a continuous decoupling of the interior
produced by an accumulation of pinned vortices in reservoirs (Alpar et
al. 1996; Ho \& Andersson 2012)\nocite{accp96,ha12}, eventually
locking up a substantial fraction of the superfluid. Such a process
would eventually be limited by the available superfluid, which is
usually considered to be the superfluid in the neutron-rich inner
crust. It seems unlikely that the process could persist long
enough to explain the low braking index of older pulsars such as the
Vela pulsar.

Several authors, notably \citet{br88,crz98,ah97,lyn04,elk+11}, have
remarked that a low braking index might be due to an increasing
dipolar magnetic field, or to an increasing inclination angle
$\alpha$. Evidence for a secular increase in $\alpha$ has been
presented by \citet{lgw+13}.  From observations of the profile of the
radio pulse over 22 years, they find an increase in the component
separation amounting to $6\times10^{-3}$ degrees per year. They remark
that this can be attributed to a similar rate of increase in $\alpha$,
which for $\alpha$ = 45 deg would account for the low value of the
braking index.  Using the long-term value of $n_{obs}=2.35$, they
found that the changing $\alpha$ gives a value of
$\dot\kappa/\kappa=2.6\times10^{-4}$ yr$^{-1}$, which could account
for the observed rate of change.

However, although the secular change in $\alpha$ may well be
sufficient to account for the whole of the reduction in the value of
the braking index below 3, it should be noted that that the
determination of $\dot\alpha$ from the observation of the separation
of the components of the pulse profile is model dependent.  Hence the
actual contribution is uncertain and it is also likely that the
relative contributions of the different processes discussed above may
be evolving with time.  Further possibilities are models of
pulsar magnetospheres which recognise an inner region corotating with
the star and which can lead to $n<3$ when the evolution of this
region's size changes at a different rate to the constantly-growing
light cylinder region \citep{mel97,cs06b,bta+06}.

\section{Long-term evolution and the $P-\dot{P}$ diagram}

\begin{figure} 
\includegraphics[width=8.5cm]{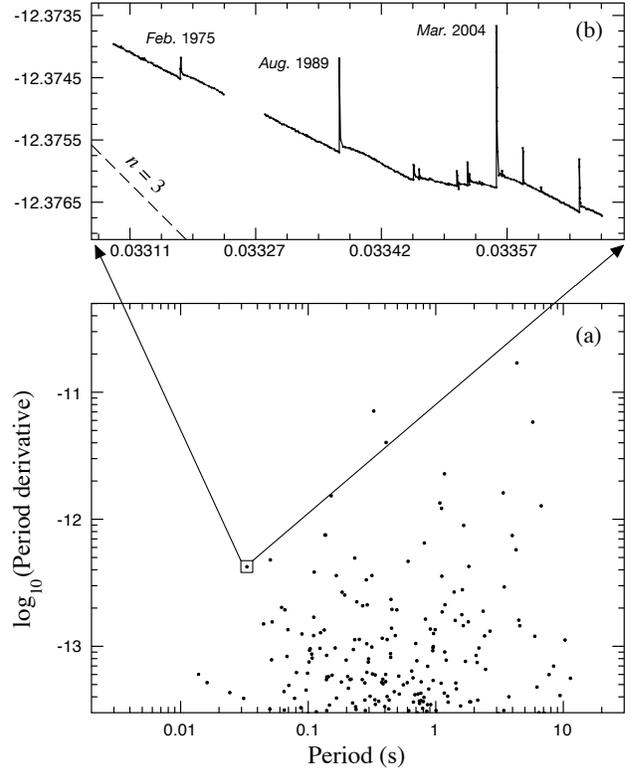}
\caption{The progress of the Crab Pulsar across the $P-\dot{P}$
diagram. a) The upper part of the standard $P-\dot{P}$ diagram and
b) an expanded view of the region containing the Crab Pulsar, showing
its motion during the past 45 years. The dot-dash line represents the
path that would be followed by a pulsar having braking index $n=3$.}
\label{ppdot}
\end{figure}

The departure from simple magnetic dipole slowdown is illustrated
dramatically in Fig.~\ref{ppdot}.  This shows a section of the
familiar $P-\dot{P}$ diagram, a log-log plot on which it is customary
to plot all known pulsars.  As a pulsar ages, it moves from left to
right across the diagram, following a path whose slope $2-n$ depends on the
braking index.  For $n=3$, the slope is $-$1.  For the Crab pulsar, for
which the mean value of $n=2.349$, the slope is $\sim-0.35$,
representing the general movement of the Crab pulsar on the diagram.
However, between glitches the evolution is with a slope of $-0.5$,
corresponding to $n=2.5$.  If the wind model is correct, the monopole
term will eventually dominate and the path across the $P-\dot{P}$
diagram will turn upwards towards a slope of +1.  As pointed out by
\citet{ac04b}, this leads into a region of the diagram where no pulsars
have been detected; however, it does lead towards the magnetars
\citep{elk+11}.  \citet{hck99} remark on the possibility that the high
slowdown rate of the magnetars may be accounted for largely or
completely by monopolar wind torque, in contrast to the current
interpretation in terms of a very high dipole magnetic field.

\section{Conclusions}

The slowdown of the rotation rate of the Crab pulsar, including the
effect of the glitches, could be described by a power law with braking
index of around $2.35$.  The spin evolution is, however, affected by
the steps in slowdown rate at glitches; removing these reveals an
underlying simple slowdown with braking index 2.519(2). The average
effect of the glitches is to increase the rate of slowdown by about
6\%.  This description of the slowdown fails during a 11-year period
which coincides with a period of increased glitch activity. 

Including the exponential component of the persistent step in
slowdown rate at glitches allows an almost complete separation of the effects
of glitches from the underlying slowdown.  This component is in the
form of an exponential with time constant 320 days, asymptotically
reaching a value which nearly doubles the previously known effect. 

The $n=2.5$ braking index, lower than the conventional value $n=3$ 
for a rotating magnetic dipole, may be attributed to a combination of
a secular increase in magnetic inclination angle and a monopolar wind torque.
If the rate of mass loss in the wind persists, the braking index is 
expected to reduce towards $n=1$, possibly accounting for the low 
index observed in the Vela pulsar.

The observed pattern of glitch activity, including the 11-year period
of increased glitch activity, does not agree well with the random
behaviour expected from self-organised criticality.

\section*{Acknowledgements}
Pulsar research at JBCA is supported by a Consolidated Grant from the UK 
Science and Technology Facilities Council (STFC). 
C.M.E. acknowledges the support from STFC and FONDECYT (postdoctorado 3130512).


\label{lastpage}
\end{document}